# Picosecond all-optical switching of magnetic tunnel junctions

Jun-Yang Chen†, Li He†, Jian-Ping Wang* and Mo Li*

*Department of Electrical and Computer Engineering, University of Minnesota,*

*Minneapolis, MN 55455, USA.*

†These authors have equal contributions to this work.

*Corresponding authors. Email: **moli@umn.edu** and **jpwang@umn.edu**.

**Control of magnetism without using magnetic fields enables large-scale integration of spintronic devices for memory, computation and communication in the beyond-CMOS era[1-7]. Mechanisms including spin torque transfer[8], spin Hall effect[9], and electric field or strain assisted switching[10,11] have been implemented to switch magnetization in various spintronic devices. Their operation speed, however, is fundamentally limited by the spin precession time to be longer than 10-100 picoseconds[12-14]. Overcoming such a speed constraint is critical for the prospective development of spintronics. Here we report the demonstration of picosecond all-optical switching of a magnetic tunnel junction (MTJ)[15-17]—the building block of spintronic logic and memory—only using single telecom-band, infrared laser pulses. This first optically switchable MTJ uses ferrimagnetic GdFeCo as the free layer, and its switching is directly readout by measuring its tunneling magnetoresistance with a ΔR/R ratio of 0.6%. An instrument limited switching repetition rate at MHz has been demonstrated, but the fundamental limit should be higher than tens of GHz. This result represents an important step toward integrated opto-spintronic devices that combines spintronics and photonics**

**technologies to enable ultrafast conversion between fundamental information carriers of electron spins and photons.**

Ultrafast optical manipulation of the magnetism in magnetic materials is both fundamentally intriguing and practically important for it can achieve magnetization reversal without the application of a magnetic field and faster than the spin precession time, which sets the operation speed limit of most spintronic devices[18,19]. More recently, all-optical switching with sub-picosecond laser pulses has been achieved in a plethora of material systems ranging from ferrimagnetic alloys of rare-earth elements and transition metals (RE-TM) such as GdFeCo, TbCo, and TbFe [20-24], to synthetic ferrimagnetic multilayers of RE-TM[25], and ferromagnetic multilayers of Co/Pt and Co/Ni [26]. Among those, the GdFeCo system stands out as its magnetization reversal can be achieved with single laser pulses as short as a few tens of femtoseconds[27,28]. In contrast, switching of other materials is through the cumulative heating effect of multiple laser pulses and thus much slower[24]. Briefly, optical switching of GdFeCo is achieved by ultrafast heating of the material by laser pulses to a highly non-equilibrium state. Because the Gd sublattice demagnetizes slower than the Fe,Co sublattice, the anti-ferromagnetically coupled system forms a transient ferromagnetic state within sub-picosecond and relaxes spontaneously to the reversed magnetization state after cooling down[28].

All-optical switching (AOS) has been extensively explored for its prospects of enabling ultrafast magnetic recording and operation of spintronic devices. Nevertheless, there has not been any demonstration of AOS in realistic spintronic devices, such as magnetic tunnel junctions (MTJs) [15-17]. Since previous studies of AOS have only been performed on a single magnetic layer, how additional magnetic layers affect the AOS phenomenon remains unknown. In this work, we develop a perpendicularly magnetized MTJ (p-MTJ) employing GdFeCo as the free layer. We demonstrate, for the first time, all-optical switching of an MTJ, without using any external magnetic field, but rather with single sub-picosecond infrared laser pulses. The switching is read out electrically through measuring the tunneling magnetoresistance (TMR). We further demonstrate MHz repetition rate of switching GdFeCo film, which is limited by our instruments, but the fundamental upper limit of this rate should be higher than tens of GHz as has been revealed by previous time-resolved studies[28]. The demonstrated picosecond switching time of an MTJ by AOS is two orders of magnitude faster than that of any other switching methods.

The GdFeCo layer used in this study was sputtered on silicon wafers coated with thermally grown $SiO_2$. As shown in Fig. 1a, the samples have a stack structure of Ta(4 nm)/$Gd_x(Fe_{90}Co_{10})_{100-x}$(20 nm)/Ta(4 nm). The samples with Gd composition ($x$) in the range from 21% to 27% show good perpendicular magnetic anisotropy (PMA). A telecom-band infrared fiber laser, which outputs pulses with a wavelength centered at 1.55 μm and pulse width of 400 femtoseconds, was used to demonstrate AOS. The laser spot diameter of 20 μm with an optical fluence of 5.8 mJ/cm$^2$ was used consistently throughout this work. Fig. 1b shows a representative magneto-optical Kerr effect (MOKE) image of a $Gd_{26}(Fe_{90}Co_{10})_{74}$ sample after scanning the laser spot across its surface. With a low repetition rate and high scanning speed, there is no overlap between adjacent pulses. It can be seen that individual bubble domains in size similar to the laser spot have been created through AOS in domains with both up and down magnetization. AOS in our materials is independent of the laser polarization, and the laser pulses always reverse the magnetization[22].

We have systematically characterized GdFeCo samples with various compositions. The coercivity ($H_c$) and saturated magnetization ($M_s$) of the samples were measured and are summarized in Fig. 1c. We found the magnetization compensation temperature $T_M$ is at room temperature (300 K) when the Gd compensation is around 24%. AOS was observed in samples with Gd composition in the range from 22% to 26% (the purple shaded region in Fig. 1c), consistent with previous reports by several other groups[20,22,25,29].

We then demonstrate the magnetoelectric response of the GdFeCo films to AOS by measuring the anomalous Hall Effect (AHE). This method was previously developed by our group and others[30,31]. We patterned the film into pillars and deposited transparent electrodes made of 110 nm thick indium tin oxide (ITO) to measure the Hall resistance. Fig. 2a shows an optical image of a typical device with a pillar diameter of 15 μm. Using an external perpendicular magnetic field, the hysteresis loop of the anomalous Hall resistance $R_{AHE}$ was measured (Fig. 2b). The asymmetric $R_{AHE}$ for the up and down magnetized states is attributed to the slight asymmetry in the electrodes' positions. The observed rectangular hysteresis loop shows that this device has nearly 100% remanence. We next used this Hall device to demonstrate direct electrical readout of AOS. The pillar was exposed to a train of single laser pulses with a repetition rate of 0.5 Hz, and the Hall resistance was measured in real-time. The result presented in Fig. 2c shows that the Hall resistance is reversed by every laser pulse. Comparing Fig. 2b and 2c, it can be seen the values of $R_{AHE}$ reversed by AOS in Fig. 2c are exactly the same as in Fig. 2b. This result indicates that every single laser pulse completely reverses the magnetization of the GdFeCo pillar.

The above results confirm that robust AOS and magnetoelectric readout can be achieved in the GdFeCo films. Using GdFeCo as the free layer in an MTJ can enable the MTJ device to be switched by light only. To this end, we designed and fabricated a MTJ stack in the configuration of Ta(5 nm)/Pd(10 nm)/[Co(0.6 nm)/Pd(1.5 nm)]$_{\times 4}$/Co(0.8 nm)/MgO(1.8 nm)/GdFeCo(20 nm)/Ta(4 nm), as illustrated in Fig. 3a. The MgO layer is the tunneling barrier. The Ta and Pd layers are used as the buffer and the bottom electrodes. The Co/Pd multilayers are the fixed layers, which are optimized to obtain good PMA. Especially, ITO was used as a transparent top electrode to allow optical access. Fig. 3b shows an optical image of a representative MTJ device with a pillar diameter of 12 μm.

We then measured the tunneling magnetoresistance $R_{TMR}$ of the MTJ by sweeping a perpendicular magnetic field in the range of ±2 kOe. A clear TMR minor loop showing the low and high resistance states of the MTJ was measured, as shown in Fig. 3e. Because the bottom Co/Pd multilayers have a lower coercivity $H_c$ than the top GdFeCo layer, they were switched by the magnetic field while the magnetization of the GdFeCo layer remained fixed. The MTJ in the low and high resistance state has a resistance of 98.0 Ω and 98.6 Ω, respectively. Therefore, the TMR ratio $\Delta R_{TMR}/R_{TMR}$ is ~ 0.6%. We attribute the low TMR ratio to the low quality of the MgO layer because no post-deposition annealing was performed, and to the oxidation at the MgO/GdFeCo interface[32,33]. The relatively high noise in the TMR measurement is attributed to the poor interface between ITO layer with high resistivity and the Ta/GdFeCo layers (See Supplementary Material for results from devices using gold top contact showing much lower noise). Nevertheless, the device provides sufficient TMR signal to noise ratio to discern optical switching of the MTJ. The AOS measurement on the MTJ was performed in a similar way to the Hall device by setting the laser repetition rate at 0.5 Hz and monitoring the TMR value with an averaging time constant of 100 ms to obtain the result shown in Fig. 3f. It clearly shows that each laser pulse switches the MTJ between high and low resistance states. In contrast to switching using a magnetic field, the laser pulses switch the magnetization of the GdFeCo layer that is on the top of the MTJ while the magnetization of the Co/Pd multilayers at the bottom remains fixed, as evident in the MOKE images in Figs. 3c and d. The change of the TMR by AOS is 0.6 ± 0.05 Ω, the same as the value in the magnetic field measurement (Fig. 3E). We note that, in Figs. 3e and 3f, the difference in the absolute resistance values is due to the changed probe to device contact resistance in the two different measurement setup. The consistent change of the TMR value unambiguously confirms that the GdFeCo layer in the MTJ has been completely switched via AOS. The TMR ratio in this device can be increased by exchange coupling the GdFeCo layer with a ferromagnetic layer, such as CoFe and CoFeB[34,35], and by annealing to improve the quality of the MgO layer and the interfaces. Our result represents the first demonstration of all-optical switching of a realistic spintronic device by using ultrafast laser pulses. The demonstrated picosecond time scale of the switching sets a new record that is at least two orders of magnitude faster than other switching mechanisms[12-14].

We next demonstrated the repeatability of AOS using a GdFeCo Hall device similar to the one in Fig. 2. The device was exposed to trains of multiple pulses with a time spacing of 1μs, generated with a pulse picker. Our laser's base repetition rate is 1MHz so shorter time spacing is not possible. Fig. 4 shows that, when the device was exposed to two consecutive pulses, the second pulse, 1 μs after the first pulse, switched the device again and thus reset the device to its original state. As a result, no change of $R_{AHE}$ was observed because a measurement time constant of 20 ms was used. Similarly, when the device was exposed to three and four consecutive pulses, the third pulse switched the device again, so the change of $R_{AHE}$ resumed, and the fourth pulse resets the device. This result demonstrates 1 MHz AOS repetition rate of a GdFeCo Hall device. However, the fundamental switching rate of GdFeCo should be much higher. X-ray magnetic circular dichroism and time-resolved MOKE studies of GdFeCo system all revealed that as soon as a few picoseconds after the laser pulse, the magnetization of both Fe and Gd sublattices have been reversed[28]. Although the system takes more than tens of picoseconds to relax to equilibrium[36], it is possible that the subsequent switching can be performed sooner than the system reaches equilibrium. Therefore, the ultimate switching rate of AOS devices can be higher than tens of GHz, which can be investigated with time-resolved measurement in the future work.

In conclusion, we have integrated an optically switchable GdFeCo film into an MTJ device and have demonstrated ultrafast all-optical switching of this spintronic device using sub-picosecond laser pulses. The picosecond time scale of optical switching is two orders of magnitude faster than other switching methods. The use of telecom-band infrared laser sources also makes such devices compatible for integration with silicon photonics and fiber optics. Future work to improve the TMR ratio of the optically switchable MTJ and reduce the required optical fluence for switching is necessary to enable large-scale integration and practical applications. The energy per pulse required for AOS scales inversely with the device area and the efficiency of AOS also improves with reduced device area[37]. Therefore, for an AOS device with sub-wavelength dimensions[38,39], it is possible to reduce the switching pulse energy to femto-joule level. Nevertheless, the present results pave a path toward a new category of opto-spintronic devices which can directly convert ultrafast optical signals into non-volatile magnetic states of spintronic devices and thus may find novel applications that combine photonic and magnetic technologies.

FIGURES

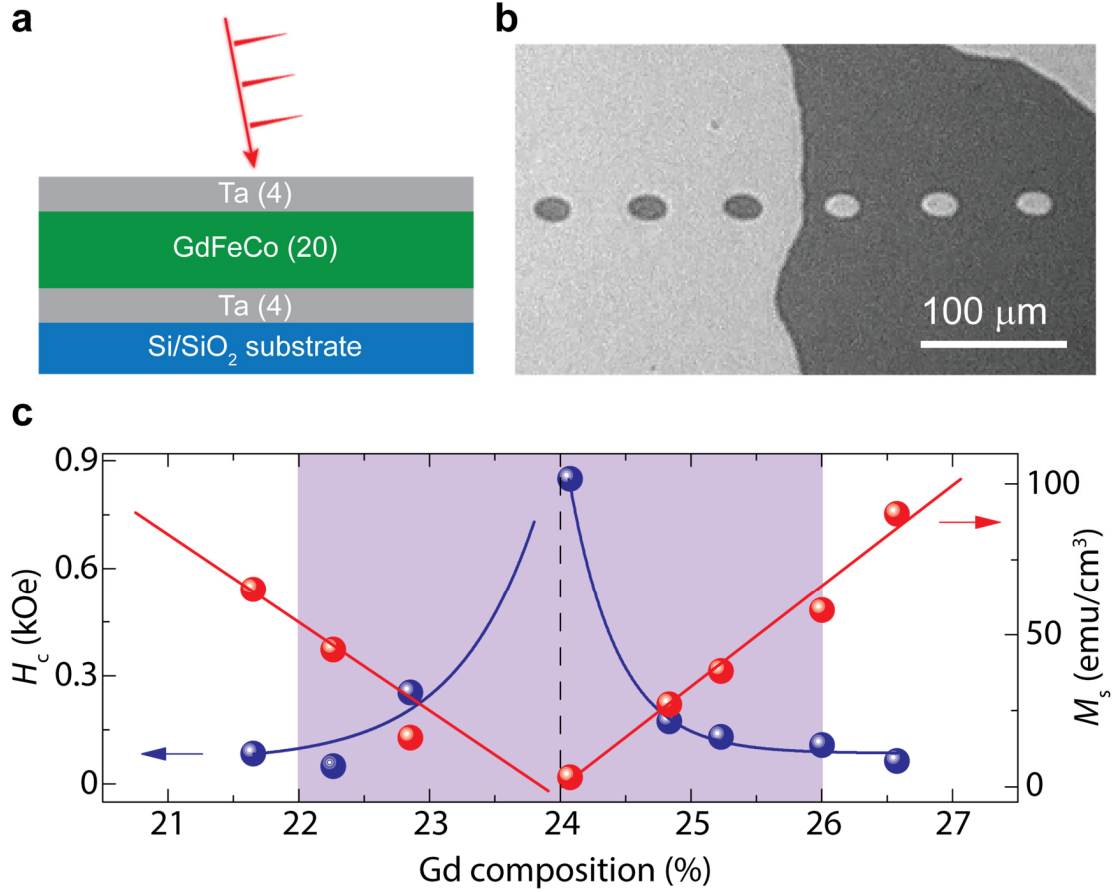

**Fig. 1. All-optical switching of GdFeCo films. a**, The schematic diagram of the GdFeCo film structures. Tantalum layers are used as buffer and capping to prevent film oxidation. **b**, The MOKE image of single bubble domains created via AOS by scanning single sub-picosecond laser pulses across two large magnetic domains with up (bright) and down (dark) magnetization. **c**, The coercivity $H_c$ (blue symbols) and saturated magnetization $M_s$ (red symbols) of GdFeCo samples versus their Gd composition ($x_{Gd}$). Samples with Gd composition in the purple shadowed region (22-26%) show AOS behavior. The solid lines are used to guide the eyes.

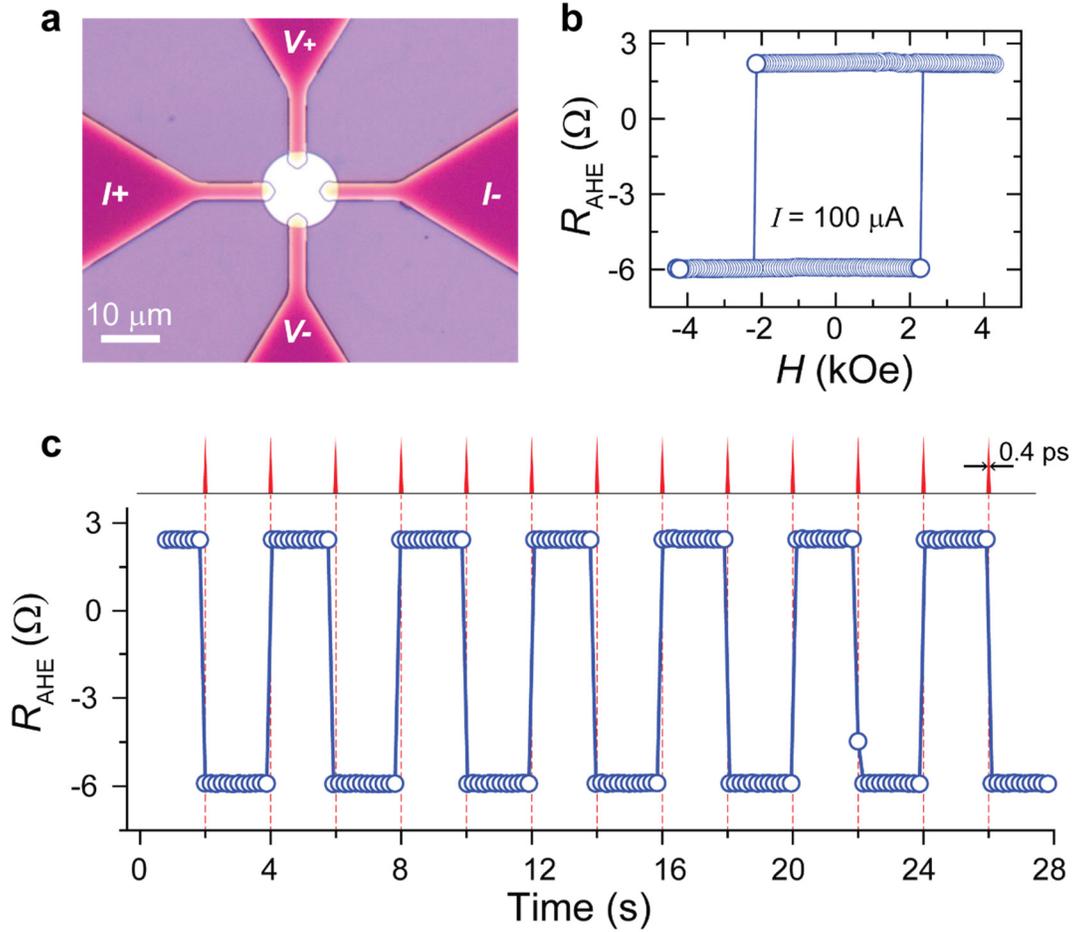

**Fig. 2. Direct magnetoelectric readout of AOS in GdFeCo film by anomalous Hall Effect (AHE). a**, The optical microscope image of a typical Hall device. The GdFeCo pillar (white) has a diameter of 15 μm, and the transparent electrodes (purple) are made of ITO/Cu. **b**, $R_{AHE}$ hysteresis loop of the device measured by sweeping a perpendicular magnetic field and applying a constant DC bias current of 100 μA. **c**, $R_{AHE}$ of the device measured in real time during AOS by 0.4 ps single laser pulses with 0.5 Hz repetition rate. The consistent change of $R_{AHE}$ in b and c confirms that the GdFeCo pillar is completed switched by the laser pulses.

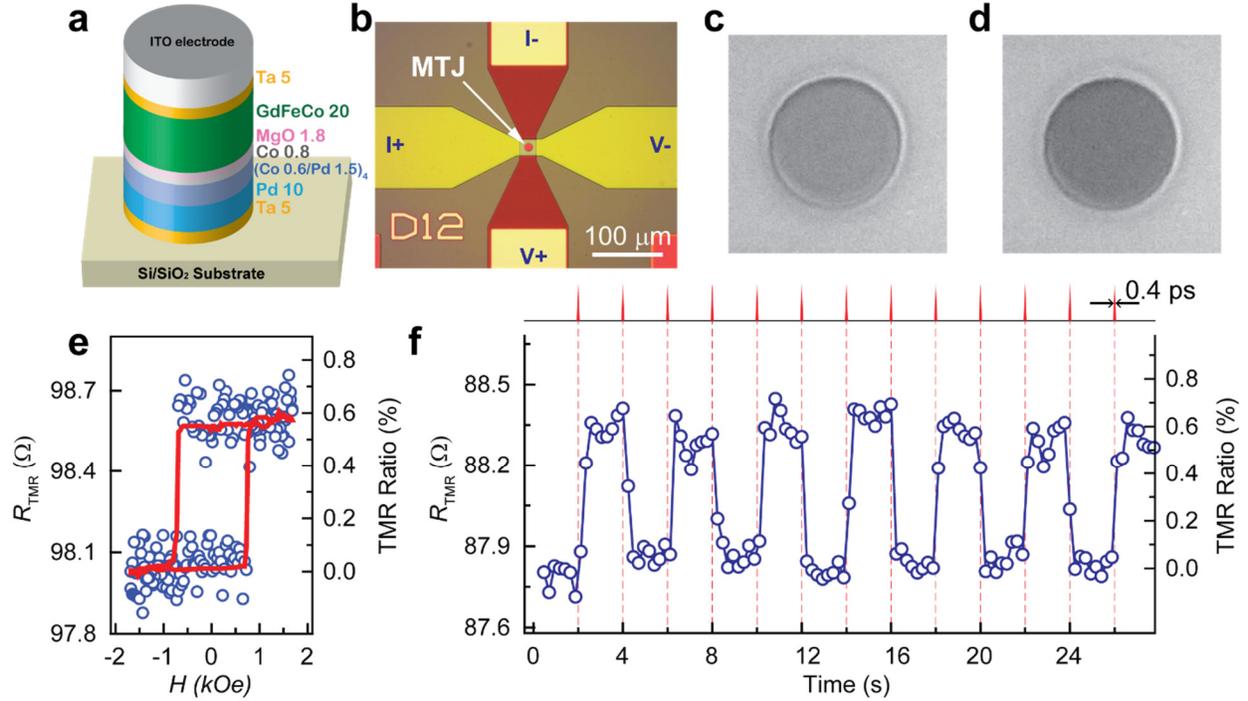

**Fig. 3. All-optical switching of an MTJ with sub-picosecond single laser pulses and without external magnetic field. a**, The schematic diagram of the MTJ structure used in the experiment. **b**, The optical microscope image of a representative MTJ device with an ITO electrode on the top for TMR measurement. (**c** and **d**) The MOKE images of the MTJ pillar before and after AOS by a single laser pulse, showing the GdFeCo layer (within the circule) is completely switched. The pillar diameter is 12 μm. **e**, The $R_{TMR}(H)$ minor loop measured by sweeping a perpendicular magnetic field, which switches the Co/Pd layers. The red line is the smoothing of the raw data (open circles). **f**, $R_{TMR}$ of the MTJ device measured during AOS by 0.4 ps single laser pulses at 0.5 Hz repetition rate. The changes of $R_{TMR}$ in **e** and **f** have the same value of ~0.6±0.05 Ω, indicating the GdFeCo layer has been completely switched.

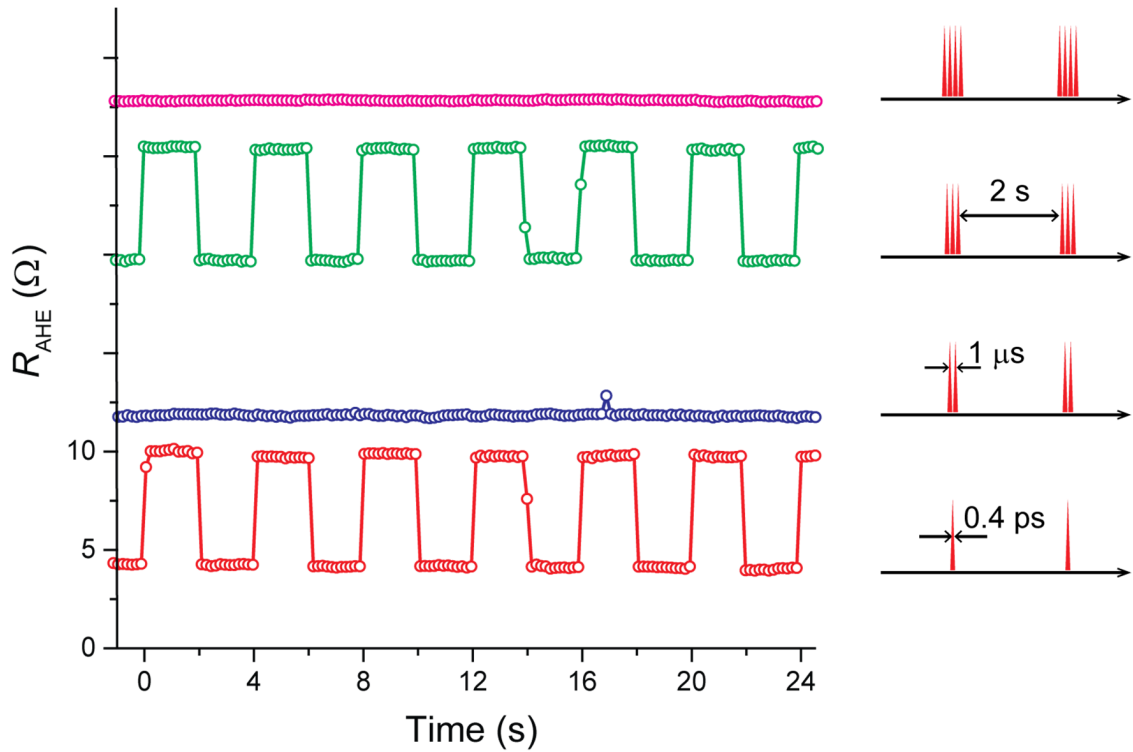

**Fig. 4. Repetitive AOS of the GdFeCo Hall Device.** The GdFeCo Hall device was exposed to trains of single (red), dual (blue), three (green) and four (magenta) consecutive pulses with 1 μs pulse-to-pulse time spacing and 0.5 Hz repetition rate. The AHE resistance ($R_{AHE}$) of the device was measured in real time. The device was switched repetitively by the pulses: the second pulse resets the switching by the first pulse and the fourth pulse resets the switching by the third pulse. Therefore, no $R_{AHE}$ change is measured for trains of double and quadruple pulses. The result demonstrates 1 MHz AOS repetition rate, limited by the instruments.

**ACKNOWLEDGMENTS**

We thank Y. C. Lau, G. Atcheson, and J. M. D. Coey at Trinity College Dublin for help on film deposition at the initial stage of the research, Steve Koester for the assistance of measurement requiring a high magnetic field, and Paul Crowell for help discussions. This work was supported by C-SPIN, one of six centers of STARnet, a Semiconductor Research Corporation program, sponsored by MARCO and DARPA. Parts of this work were carried out at the University of Minnesota Nanofabrication Center, which receives partial support from NSF through the NNIN program, and the Characterization Facility, which is a member of the NSF-funded Materials Research Facilities Network via the MRSEC program.